\documentclass[twocolumn,pre,aps,showpacs,amsmath,amssymb,superscriptaddress]{revtex4}

\usepackage{hyperref}
\usepackage{amsmath,amssymb}
\usepackage{amsfonts,amsthm}
\usepackage{graphics}
\usepackage{graphicx}
\usepackage{dcolumn}
\usepackage{color}
\usepackage{bm}

\usepackage[normalem]{ulem}

\begin{document}


\title{Anticipated synchronization in human EEG data: unidirectional causality with negative phase-lag}

\author{Francisco-Leandro P. Carlos}
\affiliation{Instituto de F\'{\i}sica, Universidade Federal de Alagoas, Macei\'{o}, Alagoas 57072-970 Brazil.}

\author{Maciel-Monteiro Ubirakitan}
\affiliation{Grupo de Neurodin\^amica, Departamento de Fisiologia e Farmacologia, Universidade Federal de Pernambuco, Recife PE 50670-901, Brazil.}
\affiliation{Spanish Foundation for Neurometrics Development, Department of Psychophysics and Psychophysiology, 30100, Murcia, Spain.}

\author{Marcelo Cairr\~ao Ara\'ujo Rodrigues}
\affiliation{Grupo de Neurodin\^amica, Departamento de Fisiologia e Farmacologia, Universidade Federal de Pernambuco, Recife PE 50670-901, Brazil.}


\author{Mois\'es Aguilar-Domingo}
\affiliation{Spanish Foundation for Neurometrics Development, Department of Psychophysics and Psychophysiology, 30100, Murcia, Spain.}
\affiliation{Department of Human Anatomy and Psychobiology, Faculty of Psychology, University of Murcia, 30100 Espinardo Campus, Murcia, Spain.}


\author{Eva Herrera-Guti\'errez}
\affiliation{Department of Developmental and Educational Psychology, Faculty of Psychology, University of Murcia, 30100 Espinardo Campus, Murcia, Spain.} 

\author{Jes\'us G\'omez-Amor}
\thanks{Deceased}
\affiliation{Department of Human Anatomy and Psychobiology, Faculty of Psychology, University of Murcia, 30100 Espinardo Campus, Murcia, Spain.}

\author{Mauro Copelli}
\affiliation{Departamento de F\'isica, Universidade Federal de Pernambuco, Recife PE 50670-901, Brazil.}

\author{Pedro V. Carelli}
\affiliation{Departamento de F\'isica, Universidade Federal de Pernambuco, Recife PE 50670-901, Brazil.}


\author{Fernanda S. Matias}
\thanks{fernanda@fis.ufal.br}
\affiliation{Instituto de F\'{\i}sica, Universidade Federal de Alagoas, Macei\'{o}, Alagoas 57072-970 Brazil.}

\begin{abstract}
Understanding the functional connectivity of the brain has become a major goal of neuroscience.
In many situations the relative phase difference, together with coherence patterns, 
have been employed to infer the direction of the information flow.
However, it has been recently shown in local field potential data from monkeys the existence of a synchronized regime 
in which unidirectionally coupled areas can present both positive and negative phase differences.
During the counterintuitive regime, called anticipated synchronization (AS), the phase difference does not reflect the causality. 
Here we investigate coherence and causality at the alpha frequency band ($f\sim10$~Hz) between pairs of  electroencephalogram (EEG) electrodes in humans during a GO/NO-GO task. 
We show that human EEG signals can exhibit anticipated synchronization, 
which is characterized by a unidirectional influence from an electrode A to an electrode B, but the electrode B leads the electrode A in time.
To the  best of our knowledge, this is the first verification of AS in EEG signals and in the human brain.
The usual delayed synchronization (DS) regime is also present between many pairs. 
DS is characterized by a unidirectional influence from an electrode A to an electrode B and a positive phase difference between A and B which indicates that the electrode A leads the electrode B in time.
Moreover we show that EEG signals exhibit diversity in the phase relations: the pairs of electrodes can present in-phase, anti-phase, or out-of-phase synchronization
with a similar distribution of positive and negative phase differences.
\end{abstract}
\maketitle

%
%

\section*{Introduction}

The extraordinary ability of humans to model 
and predict facts are one of the prerequisites 
for both action and cognition. These capacities 
emerge from the various synchronous rhythms generated 
by the brain~\cite{Buzsaki06,Wang10}, which represent 
a core mechanism for neuronal communication~\cite{Fries05}.
In particular, phase synchronization~\cite{Pikovsky01} has been 
related to selective attention~\cite{Doesburg08,Maris13},
large-scale information integration~\cite{Varela01} and 
memory processes~\cite{Fell11,Dotson14}.
Despite huge evidence of zero-lag synchronization in the brain~\cite{Wang10},
there is a growing number of studies reporting non-zero 
phase differences between synchronized brain 
areas~\cite{Maris13,Dotson14,Gregoriou09,Tiesinga10,Bastos15,Livingstone96}.
It has been assumed that phase diversity plays 
an important role in fast cognitive processes~\cite{Maris16}.


In many situations the phase, together with coherence patterns, 
have been employed to infer the direction of the information flow
~\cite{Korzeniewska03,Marsden01,Williams02,Schnitzler05,Sauseng08,Gregoriou09}.
The assumption is typically that the phase difference reflects the transmission time of
neural activity.
However, this assumed
relationship is not theoretically justified~\cite{Thatcher12}. Particularly, during special synchronized regimes,
the phase difference does not reflect the causality~\cite{Brovelli04,Salazar12,Matias14,Hahs11,Vakorin14}.

It has been shown that a monkey performing a cognitive task can present 
unidirectional influence from a cortical region A to another region B
with a negative phase difference between the two areas~\cite{Brovelli04,Salazar12,Matias14}.
This means that the receiver region B can lead the activity of A.
For example, it has been observed
that during the waiting period of a GO/NO-GO task, a macaque monkey present
unidirectional causality from the somatosensory cortex to the motor cortex with a negative phase~\cite{Brovelli04,Matias14}.
A similar apparent incongruence has been verified between PreFrontal Cortex (PFC) and Posterior Parietal Cortex (PPC)
in monkeys performing a working memory task~\cite{Salazar12}. 
The information flows from the PPC to the PFC but the activity of the PFC leads the activity of the PPC by $2.4$ to $6.5$~ms.

These experimental results have been compared to a model of
two unidirectionally coupled neuronal populations~\cite{Matias14}. 
The phase difference between the sender and the receiver population
can be controlled by the inhibitory synaptic conductance in the receiver population~\cite{Matias14} or by the amount of external noise at the receiver~\cite{DallaPorta19}. 
By construction, the information flow is always from the sender to the receiver population but the receiver can lead the sender, 
which is characterized by a negative phase difference. In other words, the sender lags behind the receiver. 
Results were corroborate using the statistical permutation quantifiers 
in the multi-scale entropy causality plane~\cite{Montani15}.
This counter-intuitive regime has been explained 
in the light of anticipated synchronization (AS) ideas~\cite{Voss00,Matias14}.

The anticipatory synchronization can be a stable solution of two dynamical systems coupled in a 
sender-receiver configuration, if the receiver is also subjected to a negative
delayed self-feedback~\cite{Voss00,Masoller01,Ciszak05,Pyragas08,Ambika09,Mayol12,Sausedo14}:
\begin{eqnarray}
\label{eq:voss}
\dot{\bf {S}} & = & {\bf f}({\bf S}(t)), \\
\dot{\bf {R}} & = & {\bf f}({\bf R}(t)) + {\bf K}[{\bf S}(t)-{\bf R}(t-t_d)]. \nonumber 
\end{eqnarray}
${\bf S}$ and ${\bf R}$ $\in \mathbb{R}^n$ are dynamical 
variables respectively representing the sender and the receiver systems.
${\bf f}$ is a vector function which defines 
each autonomous dynamical system, ${\bf K}$ is the coupling matrix and $t_d > 0$ is the delay in the receiver's negative self-feedback. 
In such system, 
${\bf R}(t) = {\bf S}(t+t_d)$ is a solution of the system, 
which can be easily verified by direct substitution 
in Eq.~\ref{eq:voss}. AS has been observed in excitable
models driven by white noise~\cite{Ciszak03}, chaotic systems ~\cite{Voss00,Pyragas08},
as well as in experimental setups with  semiconductor lasers~\cite{Sivaprakasam01,Tang03} and electronic circuits~\cite{Ciszak09}.

AS has also been observed when the self-feedback was 
replaced by parameter mismatches~\cite{Corron05,Srinivasan12,Pyragiene13,Pyragiene15,Simonov14}, inhibitory dynamical loops~\cite{Matias11,Matias14,Matias15,Pinto19}
and noise at the receiver~\cite{DallaPorta19}. It has been suggested that AS can emerge when the receiver dynamics is faster than the senders~\cite{Hayashi16,Dima18,Pinto19,DallaPorta19}.
Furthermore, unidirectionally coupled lasers reported both regimes: AS and the usual delayed
synchronization (DS, in which the sender predicts the activity of the receiver), depending on the difference between the transmission time and the feedback delay time~\cite{Liu02,Tang03}. 
The two regimes were observed to have the same stability of
the synchronization manifold in the presence of small perturbations due to noise or parameter
mismatches~\cite{Tang03}.
Neuron models can also present a transition from positive to negative phase differences (from DS to AS) depending on coupling parameters~\cite{Matias11,Matias14,Matias15,DallaPorta19}.
Therefore, the study of anticipatory regimes in biological systems (not man-made) is 
receiving more attention in the last years~\cite{Hayashi2016Anticipatory,Stepp10,Washburn19,Roman19}.

Here we employ spectral coherence and Granger causality (GC) measures to infer the direction of influence,
as well as the phase difference between electrodes of the EEG from 11 subjects.
We verify, for all subjects, the existence of coherent activity in the alpha band  ($f\sim10$~Hz)  between pairs of electrodes.
We also show that many of these pairs exhibit a unidirectional influence from one electrode to another and a phase difference that can be positive or negative.
In Sec.~\ref{results} and ~\ref{Appendix} we describe the experimental paradigm and EEG processing and analysis.
In Sec.~\ref{results}, we report our results, showing that when we consider all the unidirectionally coupled pairs we verify that there is a diversity in the phase relation:
they exhibit in-phase, anti-phase, or out-of-phase synchronization with similar distribution of positive and negative phase differences (DS and AS, respectively).
Concluding remarks and brief discussion of the significance of our findings for neuroscience are presented in Sec.~\ref{conclusion}.





\section{\label{results}Results}

\begin{figure}[h]
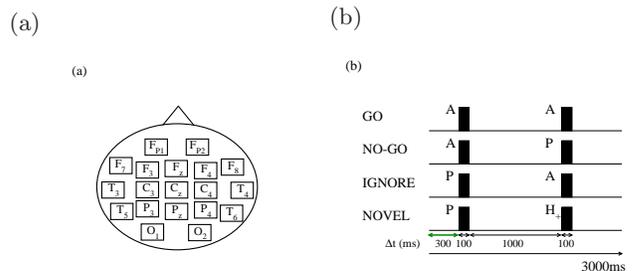

\centering
\begin{minipage}{0.48\linewidth}
\begin{flushleft}(a)%
\end{flushleft}%
\centering
\includegraphics[width=0.6\columnwidth,clip]{1acabeca.eps}
\end{minipage}
\begin{minipage}{0.48\linewidth}
\vspace{0.2cm}
\begin{flushleft}(b)%
\end{flushleft}%
\centering
\includegraphics[width=0.9\columnwidth,clip]{1btask.eps}
\end{minipage}
\caption{
{\bf Experimental paradigm.} 
(a) 10/20 System of EEG electrodes placement employed in the experiments.
(b) GO/NO-GO task based on three types of stimulus with images of animals (A), plants (P), and people (H$_+$).
After a waiting window of 300~ms, two stimulus were presented for 100 milliseconds, with a 1000 ms inter-stimulus-interval.
If both stimulus are animals (AA) the participant should press a button as quickly as
possible (see Sec.~\ref{Appendix} for more details). Here we analyzed the 300~ms before the stimulus onset.
}
\label{fig:task}
\end{figure}

The experiment consists in 400 trials of a GO/NO-GO task.
In each trial a pair of
stimuli were presented after a waiting window of $300$~ms, which is the important interval for our analysis (see the green arrow in Fig.~\ref{fig:task}(b)).
Depending on the combination of stimuli, participants should press a button or not.
Oscillatory main frequency, synchronized activity and directional influence were estimated by 
the power, coherence, phase difference and Granger causality spectra
as reported in Matias et al.~\cite{Matias14} (see more details in Sec.~\ref{Appendix}).

\begin{figure*}[!t]
\centering
\includegraphics[width=0.9\textwidth,clip]{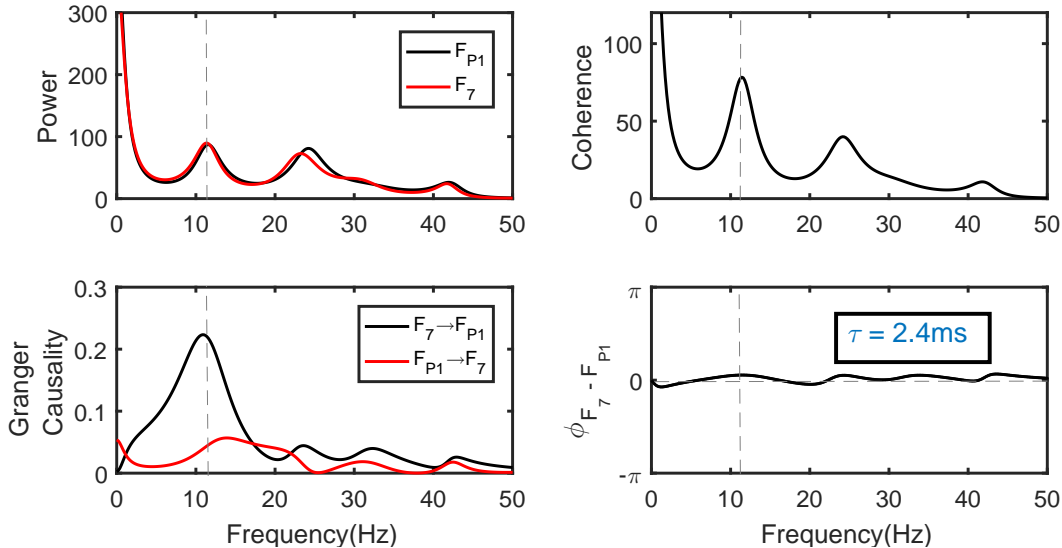}
\caption{
{\bf Unidirectional causality with positive phase-lag characterizes the delayed synchronization regime (DS).}
Power, coherence, Granger causality and phase spectra between electrodes $F_7$ and $F_{P1}$ for volunteer 439.
The pair is synchronized with main frequency 
$f_{peak}=11.4$~Hz (given by the peak of the coherence, grey dashed lines). 
The Granger causality peak around $f_{peak}$ reveals a directional influence from site $F_7$ to $F_{P1}$ and the phase difference 
at the main frequency $\Delta\Phi_{F_7-F_{P1}}(f_{peak})=0.1727$~rad shows that $F_7$ leads $F_{P1}$ (with an equivalent time delay $\tau=2.4$~ms).
}
\label{fig:GC-DS}
\end{figure*}

Synchronization between electrodes $l$ and $k$ 
can be characterized by a peak in the coherence spectrum $C_{lk}(f_{peak})$.
The phase difference $\Delta\Phi_{l-k}$ at the peak frequency 
$f_{peak}$ provides the time delay $\tau_{lk}$ between the electrodes. 
The direction of influence is given by the 
Granger causality spectrum. Whenever an electrode $l$ strongly and
asymmetrically G-causes $k$, we
refer to $l$ as the sender (S) and to $k$ as the receiver (R) and the link between $l$ and $k$ is considered 
a unidirectional coupling from $l$ to $k$ (S $\to$ R). 
After determining which electrode is the sender and which one is the receiver we analyze the sign 
of $\Delta\Phi_{S-R}$ to determine the synchronized regime. Unless otherwise stated we analyze only the unidirectionally connected pairs.

\subsection*{Delayed synchronization (DS): unidirectional causality with positive phase-lag}

Typically when a directional influence is verified from A to B, 
a positive time delay is expected, indicating that A's activity temporally precedes that of B~\cite{Gregoriou09,Sharott05}. 
This positive time delay characterizes the intuitive regime called delayed synchronization (DS, or also retarded synchronization) in which the sender is also the leader~\cite{Tang03}. 
In neuronal models the time delay between A and B can reflect the characteristic time scale of the synapses between A and B but 
can also be modulated by local properties of the receiver region B~\cite{Matias14,DallaPorta19}.

In Fig.~\ref{fig:GC-DS} we show an example of DS between the sites F$_7$ and F$_{P1}$ for volunteer 439. 
Power and coherence spectra present a peak at $f_{peak}=11.4$~Hz. 
At this frequency, the activity of F$_7$ G-causes F$_{P1}$, but not the other way around. 
The positive sign of the phase  $\Delta\Phi_{F_7-F_{P1}}(f_{peak})=0.1727$~rad indicates that the sender electrode F$_7$ 
leads the receiver electrode F$_{P1}$ with a positive time delay $\tau=2.4$~ms.

\subsection*{Anticipated synchronization (AS): unidirectional causality with negative phase-lag}

\begin{figure*}[!th]
\centering
\includegraphics[width=0.9\textwidth,clip]{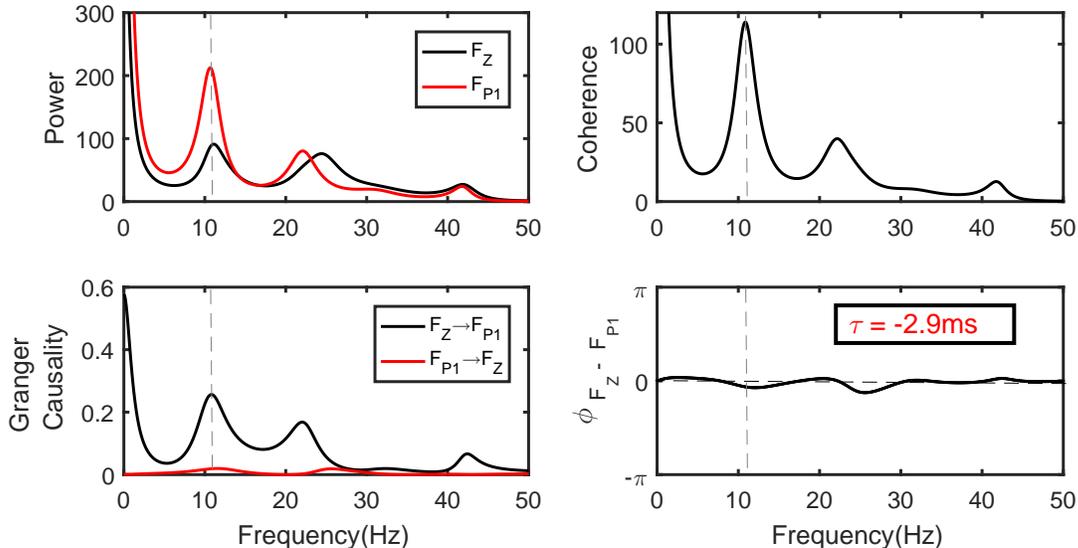}
\caption{
{\bf Unidirectional causality with negative phase-lag characterizes anticipated synchronization (AS). }
Power, coherence, Granger causality and phase spectra between sites $F_Z$ and $F_{P1}$ for volunteer 439.
The electrodes are synchronized with main frequency  $f_{peak}=10.8$~Hz (given by the peak of the coherence, grey dashed lines).
The Granger causality peak around $f_{peak}$ reveals a directional influence from site 
$F_Z$ to $F_{P1}$. $F_Z$ G-causes $F_{P1}$, but the negative phase difference 
at the main frequency $\Delta\Phi_{F_Z-F_{P1}}(f_{peak})=-0.1969$~rad (which is equivalent to a time delay $\tau=-2.9$~ms) indicates that
$F_{P1}$ leads $F_Z$ in time. 
} \label{fig:GC-AS}
\end{figure*}


Despite the fact that phase differences and coherence patterns, 
have been employed to infer the direction of the information flux
~\cite{Marsden01,Williams02,Schnitzler05,Sauseng08,Gregoriou09,Korzeniewska03},
our results imply that if we consider only the coherence and phase-lag 
we could infer the wrong direction of influence between the involved pairs. 
Such counter-intuitive regime exhibiting unidirectionally causality 
with negative phase difference has first been reported in the brain as a mismatch
between causality and the sign of the phase difference in local field potential 
of macaque monkeys during cognitive tasks~\cite{Brovelli04,Salazar12}.
Afterwards, it has been reported that the apparent paradox could be explained in the 
light of anticipated synchronization ideas~\cite{Matias14}.
Here we show that human EEG signals can also present unidirectional influence 
with negative phase-lag. As far as we know, this is the first evidence of AS in human EEG data.

An example of anticipated synchronization between EEG electrodes is shown in Fig.~\ref{fig:GC-AS}. The sites $F_Z$ and $F_{P1}$
exhibit a peak at alpha band in the power and coherence spectra for $f_{peak}=10.8$~Hz (Fig.~\ref{fig:GC-AS}) for volunteer 439. 
The Granger causality spectra presents a peak from 
$F_Z$ to $F_{P1}$ but not in the opposite direction, indicating that
$F_Z$ G-causes $F_{P1}$ at $f_{peak}=10.8$~Hz.
However, the negative sign of the angle $\Delta\Phi_{F_Z-F_{P1}}(f_{peak})=-0.1969$~rad indicates that the activity of F$_Z$ lags behind the activity of F$_{P1}$. 
The time delay associated to $\Delta\Phi_{F_Z-F_{P1}}(f_{peak})$ is $\tau=-2.9$~ms.

It is worth mentioning that for linear phase responses, which is the case for a simple monochromatic sinusoidal function, 
the phase delay and the group delay  (defined by the derivative of phase with respect to frequency) are identical. 
In this case, both phase and group delays may be interpreted as the actual time delay between the signals.
For time series that are synchronized in a broad frequency band, the group delay could be useful to estimate 
the time difference between the signals. Indeed, a negative group delay has been associated with anticipatory dynamics~\cite{Voss16,Voss16Negative,Voss18}
and it is comparable to the time difference obtained by the cross-correlation function~\cite{Voss16}.
Here, we verified that some AS pairs present both negative phase delay and negative group delay (as in the example shown in Fig.~\ref{fig:GC-AS}). 
However, this is not the case for all AS pairs in the analyzed data. We have found all possible combinations for the signs of phase and group delays for both DS and AS. 
A further investigation of the relation between phase delay and group delay in brain signals is out of the scope of this paper and should be done elsewhere.”

\subsection*{Zero-lag synchronization (ZL)}

Zero-lag (ZL) synchronization has been widely documented in experimental data since its first report in the cat visual cortex~\cite{Gray89}. 
It has been related to different cognitive functions such as perceptual integration and the execution of coordinated motor behaviours ~\cite{Roelfsema97,Varela01,Fries05,Uhlhaas09}. 
Despite many models showing that bidirectional coupling between areas promotes zero-lag synchronization~\cite{Vicente08,Gollo14}, 
it is also possible to have ZL between unidirectional connected populations~\cite{Matias14,Matias15,DallaPorta19}. 
In these systems, nonlinear properties of the receiver region can compesate characteristic synaptics delays and the two systems synchronize at zero phase.

\begin{figure*}[!th]
\centering
\includegraphics[width=0.9\textwidth,clip]{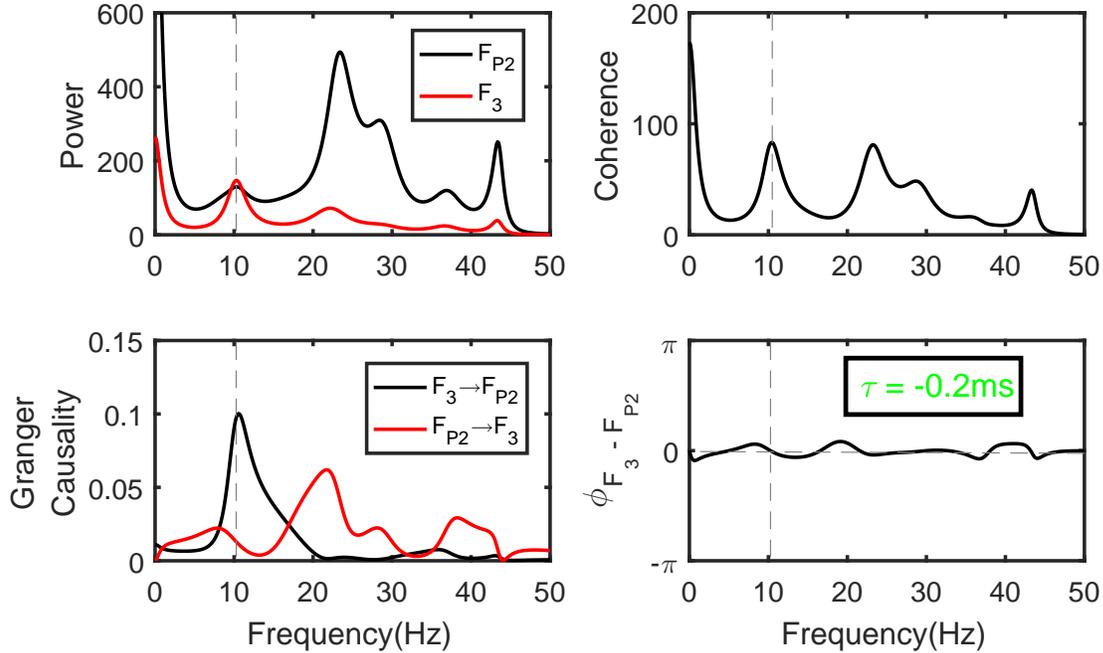}
\caption{
{\bf Unidirectional causality with zero-lag synchronization (ZL, defined by $\Delta\Phi \simeq 0$). }
Power, coherence, Granger causality and phase spectra between electrodes $F_3$ and $F_{P2}$ for volunteer 439. Sites are synchronized with main frequency (given by the peak of the coherence, brown dashed lines) 
$f_{peak}=10.4$~Hz. The Granger causality peak around $f_{peak}$ indicates that site $F_3$ unidirectionally influences $F_{P2}$. The time delay between both is almost zero $\tau=-0.2$~ms ($\Delta\Phi_{F_3-F_{P2}}(f_{peak})=-0.0164$~rad).
}
\label{fig:GC-ZL}
\end{figure*}

We consider zero-lag whenever $\arrowvert \Delta\Phi_{S-R}(f_{peak})\arrowvert < 0.1$~rad. 
In Fig.~\ref{fig:GC-ZL} we show power, coherence, Granger causality and phase spectra between electrodes $F_3$ and $F_{P2}$ for volunteer 439.
These sites are synchronized with main frequency
$f_{peak}=10.4$~Hz and $\Delta\Phi_{F_3-F_{P2}}(f_{peak})=-0.0164$~rad which provides $\tau=-0.2$~ms.

\subsection*{Anti-phase synchronization}

\begin{figure*}[!th]
    \centering
    \includegraphics[width=0.9\textwidth,clip]{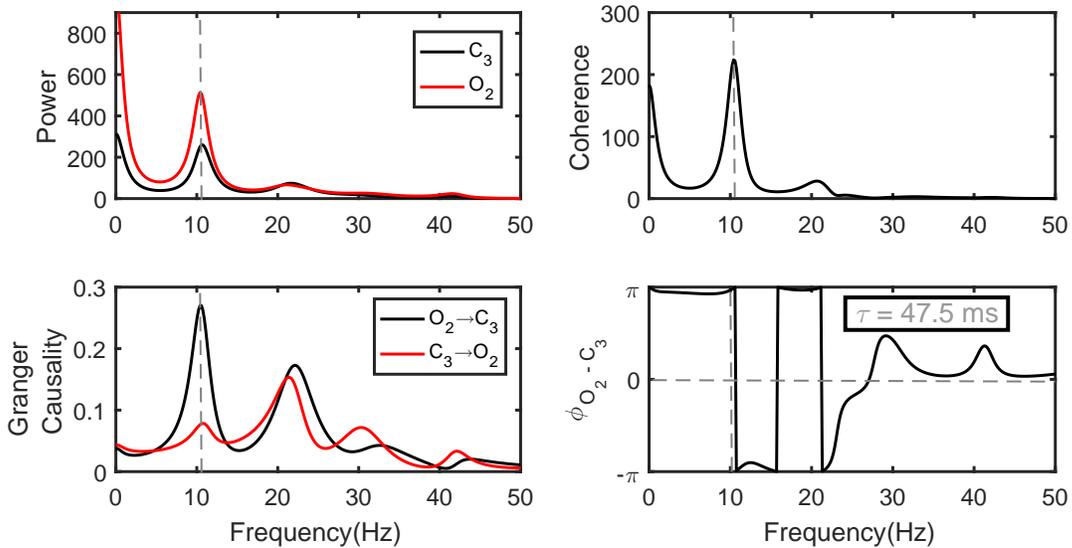}
    \caption{
    {\bf Unidirectional causality with anti-phase synchronization (AP, defined by $\Delta\Phi \simeq \pm\pi$).}
Power, coherence, Granger causality and phase spectra between electrodes $O_2$ and $C_3$ for volunteer 439.
The activity of the electrodes are synchronized with main frequency $f_{peak}=10.4$~Hz (grey dashed lines). 
The Granger causality peak around $f_{peak}$ reveals a directional influence from  $O_2$ to $C_3$ and 
the phase spectrum shows that $\Delta\Phi_{O_2-C_3}(f_{peak})=3.1031$~rad (which provides $\tau = 47.5$~ms).
}
    \label{fig:GC-AP}
\end{figure*}

Participants can also exhibit anti-phase synchronization between electrodes.
We define anti-phase synchronization (AP) when $ \pi - 0.1 < \arrowvert \Delta\Phi_{S-R}(f_{peak})\arrowvert < \pi + 0.1$~rad. 
In Fig.~\ref{fig:GC-AP} we show power, coherence, Granger causality and phase spectra between electrodes $O_2$ and $C_3$ for volunteer 439. The site $O_2$ G-causes $C_3$ 
and the time delay between them is $\tau= 47.5$~ms which is almost half of a period for the $f_{peak}=10.4$~Hz.

\begin{table}[!ht]
\centering
\caption{
{\bf Number of unidirectionally connected pairs for all subjects together:} 
separated by phase-synchronization regime along the lines and by the direction of influence along the columns.
}
\begin{tabular}{|c| c| c| c| c |}
\hline
& Unidirectional & Back-to-Front & Lateral & Front-to-Back  \\ \hline
 Total & 686 & 430  &90 & 166  \\ \hline
ZL &93 & 39 &25 & 29  \\ \hline
DS(1) & 77 & 25      &14 & 38  \\ \hline
AS(1) & 99 & 51   &27 & 21  \\ \hline
AP & 174 & 135        &11 & 28  \\ \hline
DS(2) & 108  & 83    &4  & 21  \\ \hline
AS(2) & 135 & 97      &9  & 29  \\ \hline
\end{tabular}
\label{tab:totalnumbers}
 \end{table}

%

\subsection*{Phase relation diversity across pairs and subjects}

\begin{figure}[ht]
    \centering
    \includegraphics[width=0.99\columnwidth,clip]{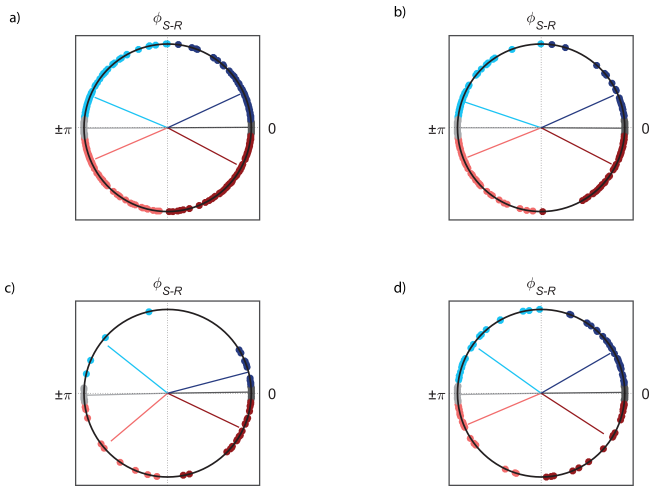}
    \caption{
    {\bf Circular phase differences distribution.} 
    The pairs are separated into six groups relative to their phase-synchronization regime: 
    zero-lag (ZL, dark gray),  anti-phase (AP, light gray), delayed synchronization in the first quadrant (DS(1),dark blue),
    delayed synchronization in the second quadrant (DS(2), light blue),
    anticipated synchronization in the fourth quadrant (AS(1), dark red), 
    anticipated synchronization in the third quadrant (AS(2), light red). 
    (a) Phase of all 686 unidirectionally connected pairs: (b) 430 pairs showing back-to-front influence, (c) 90 pairs within lateral flux, (d) 166 pairs presenting front-to-back influence.
    }
    \label{fig:all_flux}
\end{figure}

Reliable phase relation diversity is a general property of brain oscillations.
It has been reported on
multiple spatial scales, ranging from very small spatial scale (inter-electrode distance $<900$ mm) in macaque~\cite{Maris13,Dotson14},
to a large spatial scale (using magnetoencephalography) in humans~\cite{Van15}.
However, the functional significance of phase relations in neuronal signals is not well defined. 
It has been hypothesized that it may support effective neuronal communication by enhancing neuronal selectivity and promoting segregation of multiple information streams~\cite{Maris16}.

\begin{figure}[ht]
    \centering
    \includegraphics[width=0.8\columnwidth,clip]{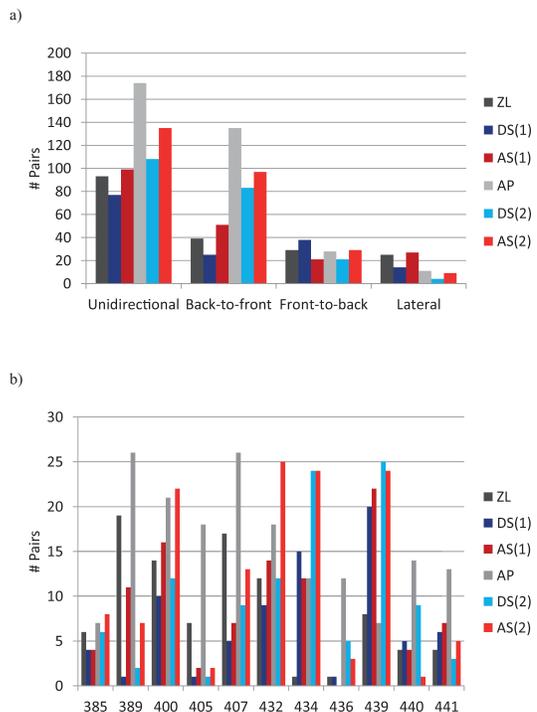}
    \caption{
    {\bf Histograms for number of pairs in each synchronized regime.}
 The colors indicate phase-synchronization regime. (a) Electrode pairs are separated by direction of influence: all unidirectional pairs, back-to-front influence, front-to-back and lateral direction. 
 (b) All unidirectional pairs separated per volunteer. 
}
    \label{fig:hist}
\end{figure}

Considering the 19 electrodes per subject, the number of analyzed pairs is 171 for each volunteer which corresponds to 1881 pairs in total. 
Among these pairs, 1394 presented a peak in the coherence spectrum at the alpha band. Regarding the Granger causality spectra, 686 pairs presented an unidirectional 
influence and 358 a bidirectional influence. 
In Fig.~\ref{fig:all_flux}(a) we show the phase-difference distribution of all 686 unidirectionally connected pairs for all volunteers in a circular plot. 
In Figs.~\ref{fig:all_flux}(b),(c),(d) we show all the pairs separated by the direction of influence: from the back to the front (430), 
lateral flux (90) and from the front to the back (166), respectively. The colors represent the four different synchronized regimes mentioned before: 
DS (blue for positive phase: $0.1 < \Delta\Phi_{S-R}(f_{peak}) < \pi - 0.1$~rad), AS (red for negative phase: $-\pi+0.1 <  \Delta\Phi_{S-R}(f_{peak}) < -0.1$~rad), 
ZL (dark grey for close to zero-phase: $\arrowvert \Delta\Phi_{S-R}(f_{peak})\arrowvert < 0.1$~rad) 
and AP (light grey for phase close to $\pm\pi$: $\pi - 0.1 < \arrowvert \Delta\Phi_{S-R}(f_{peak})\arrowvert < \pi + 0.1$~rad). 
We have also separated the DS and AS regimes into two different subcategories: DS(1) for phase in the first quadrant (dark blue), DS(2) for phase in the second quadrant (light blue), 
AS(1) for phase in the fourth quadrant (dark red) and AS(2) for phase in the third quadrant (light red). 
The number of pairs in each situation are shown in Table~\ref{tab:totalnumbers} and in Fig.~\ref{fig:hist}.

The total number of synchronized and unidirectionally connected pairs varies among volunteers, 
as well as the distribution of phases. All subjects present DS, AS, ZL and AP pairs (see Fig.~\ref{fig:hist}(b)). 
However, one subject does not present AS(1). All subjects present back-to-front, lateral and front-to-back influence 
and more pairs with back-to-front than front-to-back direction of influence. 
Considering only the back-to-front pairs, there are more AP than ZL synchronized regimes. 
This is also true if we compare all pairs in the second and third quadrant (AP, DS(2) and AS(2)) with the ones in the first and fourth (ZL, DS(1), AS(1)).

\begin{figure}[h]
    \centering
    \includegraphics[width=0.99\columnwidth,clip]{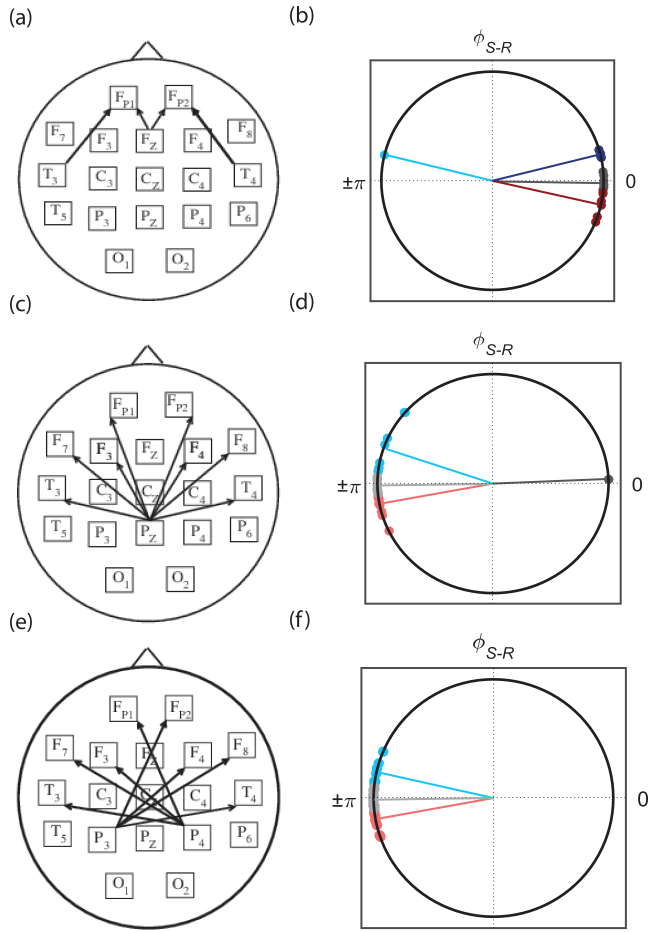}
    \caption{
    {\bf Illustrative examples of unidirectionally connected pairs and their phase relations.} 
    (a) and (b) Example of pairs with the majority of phase differences in the first and the fourth quadrants (ZL, DS(1), AS(1)): 
    $F_Z\rightarrow F_{P1}$, $F_Z \to F_{P2}$, $T_3 \rightarrow F_{P1}$ and $T_4 \rightarrow F_{P2}$. 
    (c) to (f) Example of pairs with the majority of phase differences in the second and the third quadrants 
    (AP,DS(2),AS(2)). $P_Z$, $P_{3}$ and $P_{4}$ are well connected senders. All the chosen pairs are synchronized with same direction of influence for at least 4 subjects.
    }    
    \label{fig:circjun}
\end{figure}

As illustrative examples, in Fig.~\ref{fig:circjun} we show the direction of influence between some pairs 
that have the same unidirectional back-to-front Granger for at least 4 subjects and their respective phases. 
Almost all pairs that have the electrodes $P_Z$, $P_3$ and $P_4$ as the sender present phases close to anti-phase (AP, DS(2), AS(2)), 
whereas almost all the pairs in which the sender is $F_Z$, $T_3$ or $T_4$ are synchronized close to zero-lag (ZL, DS(1), AS(1)).

Regarding back-to-front influences, no pair presented the same Granger causal relation for 9 or more subjects. 
Three pairs exhibited same unidirectional relation for 8 volunteers:  $P_Z \to F_{7}$, $P_3 \to F_{P2}$,   $O_1 \to F_{4}$; 
other 3 pairs presented the same unidirectional relation for 7 subjects:  $P_3 \to F_{4}$,  $P_3 \to F_{8}$,  $O_1 \to F_{P2}$.
Ten pairs had same Granger causal relation for 6 volunteers: $F_Z \to F_{P1}$, $P_3 \to F_{P1}$, $P_3 \to F_{3}$, $C_Z \to F_{8}$, $C_Z \to T_{3}$, $C_4 \to F_{P1}$, $C_4 \to F_{7}$, $C_4 \to F_{3}$, $O_1 \to F_{P1}$, $O_2 \to F_{4}$. 
All these 16 pairs had none or only one other subject presenting the opposite direction of the Granger causality. 
Out of these 16 pairs, only $F_Z \to F_{P1}$ is mostly synchronized close do ZL as shown in Figs.~\ref{fig:all_flux}(a) and (b), 
all others are mostly synchronized close to AP as in Figs.~\ref{fig:all_flux}(c)-(f).

\section{\label{conclusion}Conclusion}

We show that human EEG can simultaneously present unidirectional causality and diverse phase relations between electrodes.
Our findings suggest that the human brain can
operate in a dynamical regime where the information flow and relative phase-lag have opposite signs.
To the best of our knowledge this is the first evidence of unidirectional influence accompanied by negative phase differences in EEG data. 
This counter-intuitive phenomena have been previously reported as anticipated synchronization in monkey LFP~\cite{Matias14,Brovelli04,Salazar12},
in neuronal models~\cite{Ciszak03,Matias11,Pyragiene13,Sausedo14,Simonov14} and in physical systems~\cite{Sivaprakasam01,Tang03,Corron05,Ciszak09,Srinivasan12}.
Therefore, we propose that this is the first verification of anticipated synchronization in EEG signals and in human brains.

Studies estimating the actual brain connectivity using data from EEG signals should consider many relevant issues such as~\cite{Brette12book}:
the importance of common reference in EEG to estimate phase differences~\cite{Thatcher12}
and the effects of volume conduction for source localization~\cite{Nunez97,Van98}.
Our findings suggest that it is also important to take into account the possible existence of AS in connectivity studies 
and separately analyze causality and phase relations. It is worth mentioning that, it has been shown that for enough data points the 
Granger causality is able to distinguish AS and DS regimes~\cite{Hahs11}. 
However, for very well behaved time series the reconstruction of the connectivity can be confused by the phase~\cite{Vakorin14}.

Our results open important avenues for investigating how neural oscillations contribute to the neural 
implementation of cognition and behavior as well as for studying the functional significance of phase diversity~\cite{Maris13,Maris16}. 
Future works could investigate the relation between anticipated synchronzation in brain signals and anticipatory behaviors~\cite{Stepp10} such as anticipation in human-machine interaction~\cite{Washburn19}
and during synchronized rhythmic action~\cite{Roman19}.
It is also possible to explore the relation between consistent phase differences and behavioral data such as learning rate, reaction time and task performance during different cognitive tasks .
Neuronal models have shown that spike-timing dependent plasticity and the DS-AS transition together 
could determine the phase differences between cortical-like populations~\cite{Matias15}. 
However, an experimental evidence for the relation between learning and negative phase differences is still lacking.

We also suggest that our study can be potentially interesting to future researches on 
the relation between inhibitory coupling, oscillations and communication between brain areas.
On one hand, inhibition is considered to play an important role to establish the oscillatory alpha activity, 
in particular, allowing selective information processes~\cite{Klimesch07}. 
On the other hand, according to the anticipated synchronization in neuronal populations model presented in Ref.~\cite{Matias14},
a modification of the inhibitory synaptic conductance at the receiver population can modulate the phase relation between sender and receiver, 
eventually promoting a transition from DS to AS. Therefore, we suggest that the inhibition at the receiver region can control the phase difference between cortical areas, which
has been hypothesized to control the efficiency of the information exchange between these areas,
via communication through coherence~\cite{Fries05,Bastos15}.

\section{\label{Appendix}Appendix: Methods}

\subsubsection*{Subjects}
We analyzed data from 11 volunteers 
(10 women, 1 man, all right-handed) who signed to indicate informed consent to participate in the experiment.
The youngest was 32 years old and the oldest 55 years old (average 45.7 and standard deviation 7.8).
All subjects were evaluated by both psychiatrist and psychologist.
Exclusion criteria
were: perinatal problems, cranial injuries with loss of consciousness and neurological deficit,
history of seizures, medication or other drugs 24 hours before the recording, presence of psychotic
symptoms in 6 months prior the study and the presence of systemic and neurological diseases.
The experiment was not specifically designed to investigate 
the phenomena of anticipated synchronization in humans and the data analyzed here were first analyzed in Ref.~\cite{AguilarDomingo13}.
The entire experimental protocol was approved by the Commission of Bioethics of the University of Murcia (UMU, project: Subtipos electrofisiológicos y mediante estimulación eléctrica transcraneal del Trastorno por Déficit de Atención con o sin Hiperactividad).

\subsubsection*{EEG recording}

The electroencephalographic data recordings were carried out at the Spanish
Foundation for Neurometrics Development (Murcia, Spain) center using a Mitsar 201M amplifier
(Mitsar Ltd), a system of 19 channels with auricular reference. 
Data were digitized at a frequency of 250 Hz. 
The electrodes were positioned according to the international 10-20 system using
conductive paste (ECI ELECTRO-GEL). Electrode impedance was kept $<5$~K$\Omega$. 
The montage (Fig.~\ref{fig:task}(a)) include three midline sites (F$_Z$, C$_Z$ and P$_Z$) and eight sites over each hemisphere
(F$_{P1}$/F$_{P2}$,F$_{7}$/F$_{8}$,F$_{3}$/F$_{4}$,T$_{3}$/T$_{4}$,C$_{3}$/C$_{4}$,P$_{3}$/P$_{4}$,T$_{5}$/T$_{6}$ and O$_{1}$/O$_{2}$).
The acquisition was realized by WinEEG software (Version 2.92.56).
EEG epochs with excessive amplitude ($>50$~$\mu$V) were automatically
deleted. Finally, the EEG was analyzed by a specialist in neurophysiology to reject epochs with
artifacts.

\subsubsection*{Experimental task}

The EEG data were recorded while subjects performed a GO/NO-GO task (also called visual continuous performance task, VCPT).
Participants sat in an ergonomic chair 1.5 meters away
from a $17''$ plasma screen. Psytask software (Mitsar Systems) was used to present the images.
The VCPT consists of three types of stimuli: twenty images of animals (A), twenty images
of plants (P), twenty images of people of different professions (H$_+$). 
Whenever H$_+$ was presented, a $20$~ms-long artificial sound tone frequency
was simultaneously produced. The tone frequencies range from 500 to $2500$~Hz, in intervals of $500$~Hz.
All stimuli were of equal size and brightness. 

In each trial a pair of
stimuli were presented after a waiting window of $300$~ms, which is the important interval for our analysis (see the green arrow in Fig.~\ref{fig:task}(b)).
Each stimulus remains on the screen for $100$~ms, with a $1000$~ms inter-stimulus-interval.
Four different kinds of pairs of stimuli were employed: AA, AP, PP and PH$_+$. 
The entire experiment consists in 400 trials (the four kinds of pairs were randomly distributed and each one appeared 100 times).
The continuous set occurs when A is presented as the first stimulus, so the subject
needed to prepare to respond. An AA pair corresponds to a GO task and the participants are supposed to press a button as quickly as
possible. An AP pair corresponds to a 
NO-GO task and the participants should
suppress the action of pressing the button. 
The discontinuous set, in which P
is first presented, indicates that one should not respond (independently of the second stimuli). IGNORE task occurred with PP
pairs and NOVEL when PH$_+$ pairs appeared.
Participants were trained for about five minutes before beginning the experimental trials. They rested for
a few minutes when they reached the halfway point of the task. The experimental session lasted $\sim30$~min.

\subsubsection*{EEG processing and analysis}

The Power, Coherence, Granger causality and phase difference spectra
were calculated following the methodology reported
in Matias et al.~\cite{Matias14} using the auto-regressive modeling method (MVAR) implemented in the MVGC Matlab
toolbox~\cite{Barnett14}. Data were
acquired while participants were performing the GO/NO-GO visual pattern
discrimination described before. 
Our analysis focuses on $30000$ points 
representing the waiting window of $400$ trials ending with the
visual stimulus onset (green arrow in Fig.~\ref{fig:task}(b)).
This means that in each trial, the 300-ms pre-stimulus interval consists of 
75 points with a 250-Hz sample rate.

%
The preprocess of the multi-trial EEG time series consists in
detrending, demeaning and normalization of each trial. 
Respectively, it means to subtract from the time series the best-fitting line, 
the ensemble mean and divide it by the temporal standard deviation.
After these processes each single trial can be considered 
as produced from a zero-mean stochastic process.
In order to determine an optimal order for the MVAR model we obtained
the minimum of the Akaike Information Criterion (AIC)~\cite{Akaike74} 
as a function of model order. The AIC dropped
monotonically with increasing model order up to 30.

For each pair of sites $(l,k)$ we
calculated the spectral matrix element
$S_{lk}(f)$~\cite{Brovelli04,Lutkepohl93}, from which the coherence
spectrum $C_{lk}(f) = |S_{lk}|^2/[S_{ll}(f)S_{kk}(f)]$ and the phase
spectrum $\Delta\Phi_{l-k}(f) =
\tan^{-1}[\mbox{Im}(S_{lk})/\mbox{Re}(S_{lk})]$ were calculated. 
A peak of $C_{lk}(f)$ indicates synchronized oscillatory activity at the
peak frequency $f_{peak}$, with a time delay $\tau_{lk} =
\Delta\Phi_{lk}(f_{peak})/(2\pi f_{peak})$. We only consider $7<f_{peak}<13$~Hz and we use the terms time delay and phase difference interchangeably. 
It is worth mentioning that $\Delta\Phi_{l-k} = - \Delta\Phi_{k-l}$ and $-\pi <\Delta\Phi_{l-k} \leqslant \pi$. Directional influence from site
$l$ to site $k$ was assessed via the Granger causality spectrum
$I_{l\to k}(f)$~\cite{Matias14,Brovelli04,Lutkepohl93}. 
When the $I_{l\to k}(f)$ has a peak around the $f_{peak}$ obtained from the coherence spectrum, we consider that $l$ G-causes $k$. 
In order to define back-to-front, lateral or front-to-back influence we separated the electrodes in 5 lines (see Fig.~\ref{fig:task}(a)): $F_{P1}$ and $F_{P2}$;
$F_{7}$,$F_{3}$,$F_{Z}$,$F_{4}$ and $F_{8}$;
$T_{3}$,$C_{3}$, $C_{Z}$,$C_{4}$ and $T_{4}$;
$T_{5}$, $P_{3}$, $P_{Z}$, $P_{4}$, and $T_{6}$;
$O_{1}$ and $O_{2}$.

\begin{acknowledgments}
The authors thank CNPq (grants 432429/2016-6, 425329/2018-6, 301744/2018-1), CAPES (grants 88881.120309/2016-01, 23038.003382/2018-39), FACEPE (grants APQ-0642-1.05/18, APQ-0826-1.05/15), FAPEAL, UFAL and UFPE for financial support. This paper was produced as part of the activities of Research, Innovation and Dissemination Center for Neuromathematics (grant No. 2013/07699-0, S. Paulo Research Foundation FAPESP).

\end{acknowledgments}
%

\bibliography{matias}

\end{document}